\documentclass[onecollarge,natbib]{svjour2}
\usepackage[latin9]{inputenc}
\usepackage{textcomp}
\usepackage{amssymb}
\usepackage{esint}
\usepackage{graphicx}

\makeatletter
\bibpunct{[}{]}{;}{n}{}{,} % to get "[numbered]" references from natbib
\smartqed  % flush right qed marks, e.g. at end of proof
\usepackage{mathptmx}      % use Times fonts if available on your TeX system
\journalname{Few Body Systems}

\makeatother

\begin{document}

\title{Elastic and inelastic breakup of deuterons with energy below 100 MeV %\thanks{Grants or other notes
}

%\subtitle{Do you have a subtitle?\\
% If so, write it here}

\titlerunning{Deuteron breakup}        % if too long for running head

\author{B. V. Carlson \and R. Capote \and M. Sin}

\authorrunning{Carlson, Capote, Sin} % if too long for running head

\institute{B. V. Carlson \at 
    Instituto Tecnológico de Aeronáutica,
    São José dos Campos SP, Brazil \\
    \email{brett@ita.br} 
     \and 
    R. Capote \at NAPC-Nuclear Data Section,
    International Atomic Energy Agency, 1400 Vienna, Austria\\
     \and 
    M. Sin \at 
    University of Bucharest, P.O. Box MG-11, 70709
    Bucharest-Magurele, Romania 
}

\date{Received: date / Accepted: date}

\maketitle

\begin{abstract}  

We present calculations of deuteron elastic and inelastic breakup cross 
sections and angular distributions at deuteron energies below 100 MeV 
obtained using the post-form DWBA approximation. The elastic breakup 
cross section was extensively studied in the past. Very few calculations 
of inelastic breakup have been performed, however. We also analyze the 
angular momentum - energy distributions of the cross section for formation 
of the compound nucleus after inelastic breakup. 


\end{abstract}

\section{Introduction}

\label{intro} Deuteron-induced reactions are being used to produce
medical radioisotopes \cite{Betak-11} and as surrogates to other
reactions (see review \cite{Escher-12} and references therein), among
recent applications. Although they have been studied for decades \cite{Baur-76,Baur-84,Kasano-82,Austern-87},
the complexity of these reactions continues to make their theoretical
description challenging. 

The direct reaction mechanism is a major contributor to the deuteron
reaction cross section due to the particle's low binding energy.               .
Competition between elastic breakup, absorption of
only a neutron or a proton (stripping and inelastic breakup) and absorption
of the deuteron must be taken into account to determine the formation
or not of a compound nucleus and its subsequent decay. The inelastic
breakup reactions - those in which either only a neutron or a proton
is absorbed -  are particularly complex, as they form compound nuclei
with a wide range of excitation energies and angular momenta. We present
the results of a theoretical study of elastic and inelastic deuteron
breakup for a large selection of targets at incident deuteron energies
below 100 MeV. We use the zero-range post-form DWBA approximation
to calculate the elastic breakup cross section \cite{Baur-76,Baur-84}
and its extension to absorption channels to calculate the inelastic
breakup cross sections \cite{Kasano-82,Austern-87}. We discuss the
regularities and ambiguities in our results, as well as the irregularities
in the inelastic breakup energy and angular momentum distributions
that complicate their substitution by a smooth distribution obtained
from systematics.

\section{Theory}

\label{sec:1}
A reasonable theoretical description of elastic deuteron
breakup was developed and applied to a multitude of experimental data
almost forty years ago by G. Baur and collaborators \cite{Baur-76,Baur-84}.
The double-differential inelastic breakup cross section can be written
in terms of its T-matrix element as 

\begin{equation}
\frac{d^{6}\sigma^{bu}}{dk_{p}^{3}dk_{n}^{3}}=\frac{2\pi}{\hbar v_{d}}\frac{1}{(2\pi)^{6}}\left|T(\vec{k}_{p},\vec{k}_{n};\vec{k}_{d})\right|^{2}\delta(E_{d}+\varepsilon_{d}-E_{p}-E_{n}) \label{sig-bu}
\end{equation}
where $\vec{k}_{n}$ and $\vec{k}_{p}$ are the final neutron and
proton momenta, respectively, $\vec{k}_{d}$ is the initial deuteron
momentum and the sum of neutron and proton final kinetic energies
is constrained to the sum of the initial deuteron kinetic energy and
its binding energy by the $\delta$-function. The T-matrix element
can be well approximated by the post-form of the DWBA matrix element,

\begin{equation}
T(\vec{k}_{p},\vec{k}_{n};\vec{k}_{d})=\left\langle \tilde{\psi}_{p}^{(-)}(\vec{k}_{p},\vec{r}_{p})\tilde{\psi}_{n}^{(-)}(\vec{k}_{n},\vec{r}_{n})\left|v_{pn}(\vec{r})\right|\psi_{d}^{(+)}(\vec{k}_{d},\vec{R})\phi_{d}(\vec{r})\right\rangle \,,
\end{equation}
which, in turn, can be well-approximated within the zero-range DWBA
approximation

\begin{equation}
T(\vec{k}_{p},\vec{k}_{n};\vec{k}_{d})\rightarrow D_{0}\,\left\langle \tilde{\psi}_{p}^{(-)}(\vec{k}_{p},a\vec{R})\tilde{\psi}_{n}^{(-)}(\vec{k}_{n},\vec{R})\Lambda(R)\left|\right.\psi_{d}^{(+)}(\vec{k}_{d},\vec{R})\right\rangle \label{dwba-bu}
\end{equation}
by including a correction for finite-range effects $\Lambda(R)$ \cite{Buttle-64}
 and taking $D_{0}=$-125 Mev-fm$^{3/2}$.

To calculate the inelastic breakup cross section, one first analyzes
the inclusive double differential breakup cross section, here given
for the proton, again in the post-form of the DWBA. The initial state
of the target is its ground state, $\Phi_{A}$, but the
final neutron-target state, $\psi_{nA}^{c}$, can be any composite
state allowed by energy and angular momentum conservation, 

\begin{equation}
\frac{d^{3}\sigma}{dk_{p}^{3}}=\frac{2\pi}{\hbar v_{d}}\frac{1}{(2\pi)^{3}}\sum_{c}\left|\left\langle \tilde{\psi}_{p}^{(-)}(\vec{k}_{p},\vec{r}_{p})\psi_{nA}^{c}\left|v_{pn}(\vec{r})\right|\psi_{d}^{(+)}(\vec{k}_{d},\vec{R})\phi_{d}(\vec{r})\Phi_{A}\right\rangle \right|^{2}\delta(E_{d}+\varepsilon_{d}-E_{p}-E_{nA}^{c})\,.
\end{equation}
Following Kasano and Ichimura \cite{Kasano-82}, we write the $\delta$-function
as the imaginary part of an energy denominator,

\begin{eqnarray}
\frac{d^{3}\sigma}{dk_{p}^{3}} & = & -\frac{2}{\hbar v_{d}}\frac{1}{(2\pi)^{3}}\mbox{Im}\sum_{c}\left\langle \psi_{d}^{(+)}\phi_{d}\Phi_{A}\left|v_{pn}\right|\tilde{\psi}_{p}^{(-)}\psi_{nA}^{c}\right\rangle \\
 &  & \qquad\qquad\qquad\times\left(E_{d}^{+}+\varepsilon_{d}-E_{p}-E_{nA}^{c}\right)^{-1}\left\langle \tilde{\psi}_{p}^{(-)}\psi_{nA}^{c}\left|v_{pn}\right|\psi_{d}^{(+)}\phi_{d}\Phi_{A}\right\rangle \,,\nonumber
\end{eqnarray}
the target ground-state matrix element of which we interpret as a
neutron optical propagator,
\[
G_{n}^{(+)}(E_{d}+\varepsilon_{d}-E_{p})=\sum_c \left(\Phi_{A}\right.\left|\psi_{nA}^{c}\right\rangle \frac{1}{E_{d}^{+}+\varepsilon_{d}-E_{p}-E_{nA}^{c}}\left\langle \psi_{nA}^{c}\right|\left.\Phi_{A}\right)\,.
\]
This furnishes a cross section of the form

\begin{equation}
\frac{d^{3}\sigma}{dk_{p}^{3}}=-\frac{2}{\hbar v_{d}}\frac{1}{(2\pi)^{3}}\mbox{Im}\left\langle \chi_{n}(\vec{r}_{n})\left|G_{n}^{(+)}(E_{d}+\varepsilon_{d}-E_{p})\right|\chi_{n}(\vec{r}_{n})\right\rangle \,,
\end{equation}
where the effective neutron wave function is given by

\begin{equation}
\chi_{n}(\vec{r}_{n})=\left(\tilde{\psi}_{p}^{(-)}(\vec{r}_{p})\left|v_{pn}(\vec{r})\right|\psi_{d}^{(+)}(\vec{R})\phi_{d}(\vec{r})\right\rangle \,.
\end{equation}

To reduce this further, the imaginary part of the optical propagator is decomposed as 
\begin{equation}
\mbox{Im}G_{n}=\left(1+G_{n}^{\dagger}U_{n}^{\dagger}\right)\mbox{Im}G_{0}\left(1+U_{n}G_{n}\right)+G_{n}^{\dagger}W_{n}G_{n}\,.
\end{equation}

The inclusive proton emission cross section from breakup can then be separated
into an elastic and inelastic part, denoted here by $bu$ for elastic
breakup and $bf$ for inelastic breakup or breakup-fusion, corresponding to the first and second terms of the decompositon of the propagator, respectively,
\begin{equation}
\frac{d^{3}\sigma}{dk_{p}^{3}}=\frac{d^{3}\sigma^{bu}}{dk_{p}^{3}}+\frac{d^{3}\sigma^{bf}}{dk_{p}^{3}}\,.
\end{equation}

The contribution due to elastic breakup 
\begin{equation}
\frac{d^{3}\sigma^{bu}}{dk_{p}^{3}}=\frac{2\pi}{\hbar v_{d}}\frac{1}{(2\pi)^{3}}\int\frac{d^{3}k_{n}}{(2\pi)^{3}}\left|T(\vec{k}_{p},\vec{k}_{n};\vec{k}_{d})\right|^{2}\delta(E_{d}+\varepsilon_{d}-E_{p}-E_{n})\,,
\end{equation}
is just the double differential cross section of Eq. (\ref{sig-bu}) 
integrated over
the neutron momentum. The inelastic breakup cross section takes the
form of an expectation value of the imaginary part of the optical
potential, 
\begin{equation}
\frac{d^{3}\sigma^{bf}}{dk_{p}^{3}}=-\frac{2}{\hbar v_{d}}\frac{1}{(2\pi)^{3}}\left\langle \Psi_{n}(\vec{k}_{p},\vec{r}_{n};\vec{k}_{d})\right|W_{n}\left(\vec{r}_{n}\right)\left|\Psi_{n}(\vec{k}_{p},\vec{r}_{n};\vec{k}_{d})\right\rangle \,,
\end{equation}
where the effective neutron wave function is given by
\begin{equation}
\left|\Psi_{n}(\vec{k}_{p},\vec{r}_{n};\vec{k}_{d})\right\rangle =\left(\tilde{\psi}_{p}^{(-)}(\vec{k}_{p},\vec{r}_{p})G_{n}^{(+)}\left(\vec{r}_{n},\vec{r}_{n}^{\prime}\right)\left|v_{pn}(\vec{r})\right|\psi_{d}^{(+)}(\vec{k}_{d},\vec{R})\phi_{d}(\vec{r})\right\rangle \,.
\end{equation}
The physical interpretation of this cross section is simple: the deuteron
first breaks up and, after propagating further, the neutron is absorbed
while the proton is emitted. This wave function can be well-approximated
in the zero-range approximation, again including the finite range correction
$\Lambda(R)$ of Ref. \cite{Buttle-64}, as

\begin{equation}
\left|\Psi_{n}(\vec{k}_{p},\vec{r}_{n};\vec{k}_{d})\right\rangle \rightarrow D_{0}\,\left(\tilde{\psi}_{p}^{(-)}(\vec{k}_{p},a\vec{R})G_{n}^{(+)}\left(\vec{r}_{n},\vec{R}\right)\Lambda(R)\left|\right.\psi_{d}^{(+)}(\vec{k}_{d},\vec{R})\right\rangle \,.
\end{equation}

 To perform numerical calculations, the wave functions
and matrix elements are expanded in partial waves  of the orbital angular
momentum alone. We thus neglect the effects of spin-orbit coupling but
thereby reduce the number of matrix elements by a factor of almost 12.

To calculate the distorted wave functions, the Koning-Delaroche global
optical potentials \cite{Koning-03} were used in the proton and
neutron channels while the potentials of Refs. \cite{Han-06} or 
\cite{An-06} were used to describe the deuteron
scattering. Both deuteron optical potentials yield very similar results.
We used the potential of Ref. \cite{Han-06} in the calculations presented here. 

The elastic breakup matrix elements of Eq. (\ref{dwba-bu})  are only 
conditionally convergent.
A simple brute force application of standard integration methods requires
integrating to radii of the order of a nanometer. The numerical integration
can be reduced to radii of the order of picometers by using asymptotic
expansions of the Coulomb wave functions to approximate the integral
in the external region analytically. However, we have found the most
efficient means of performing the integrals to be their extension
to the complex plane as proposed by Vincent and Fortune \cite{Vincent-70}.
In this case,
we can usually limit the numerical integration to several hundreds 
of fm and could probably limit it even more, were we to
invest more effort in the evaluation of the Coulomb wave functions
in the complex plane.

\section{Calculations}

\label{sec:2}

\begin{figure*}[t]
\centering
\resizebox{18pc}{!}{\includegraphics{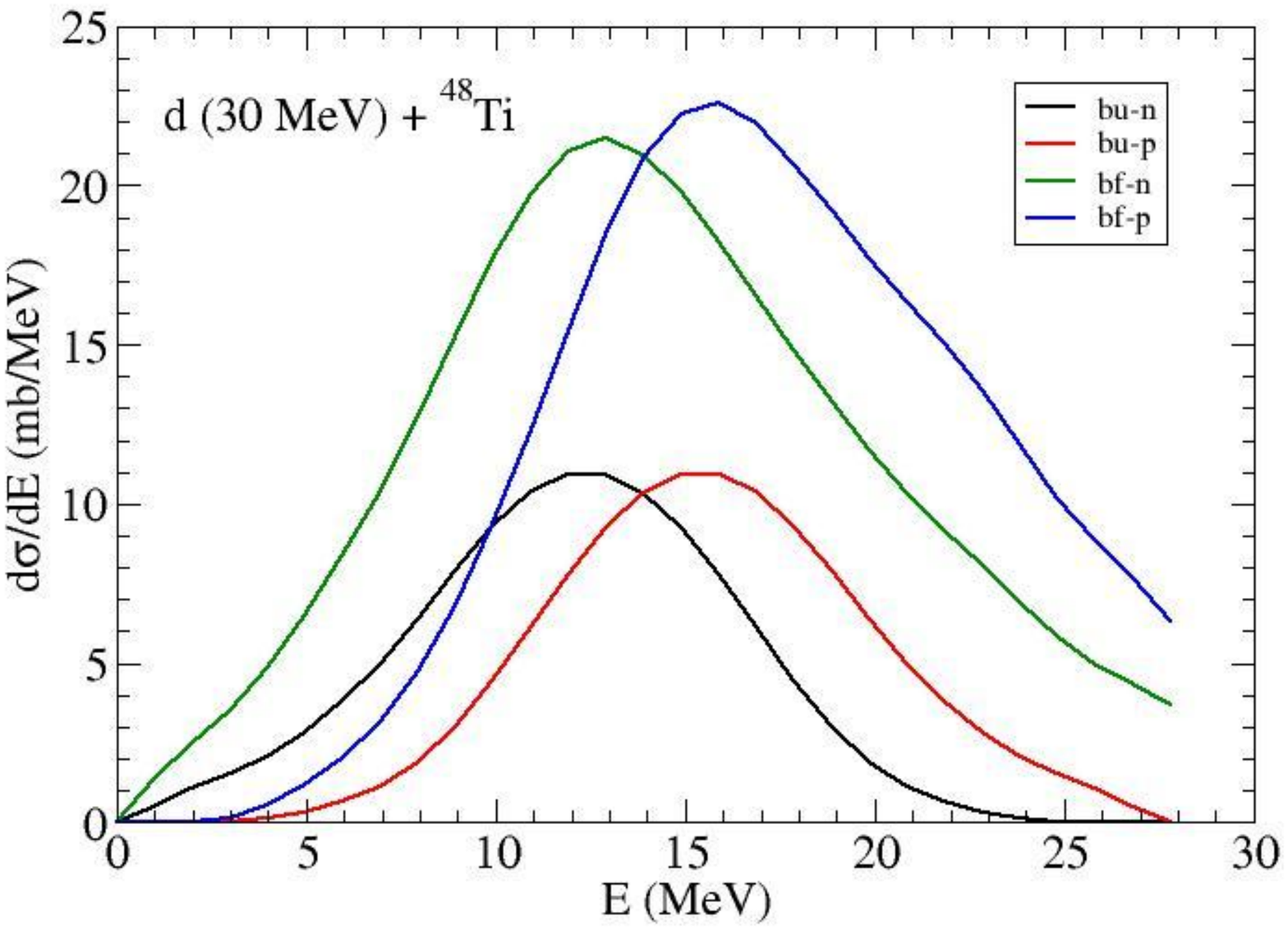}} 
\resizebox{18pc}{!}{\includegraphics{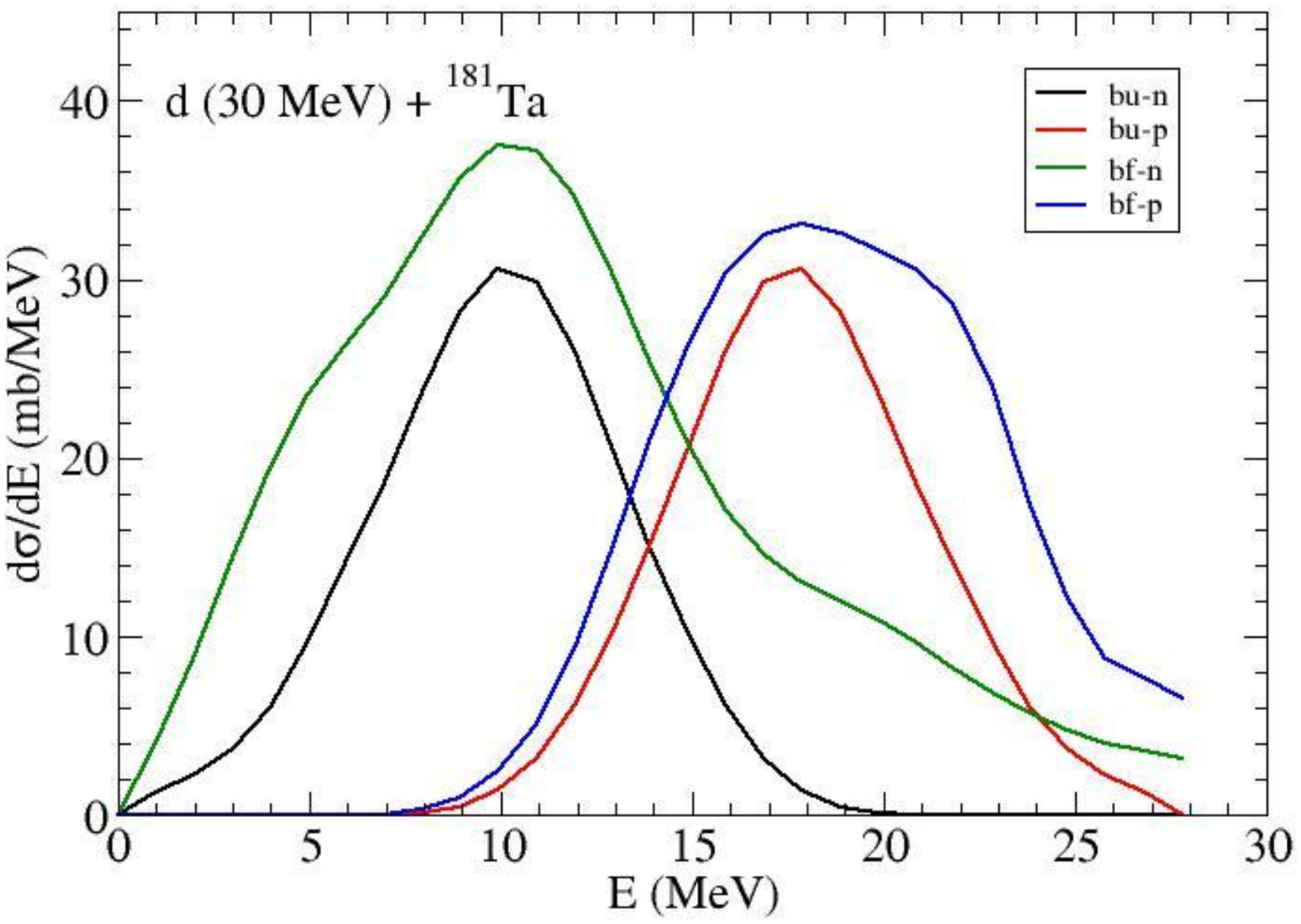}}
\caption{(a) Proton and neutron emission spectra from elastic and inelastic 
breakup spectra of 30 MeV deuterons incident on $^{48}$Ti. (b) Proton and
neutron emission spectra from elastic and inelastic breakup spectra of 30
MeV deuterons incident on $^{181}$Ta.}
\label{fig1} 
\end{figure*}

We begin by presenting calculations of neutron and proton spectra from the
elastic and inelastic breakup of 30 MeV deuterons incident on $^{48}$Ti
and $^{181}$Ta, shown in Fig. \ref{fig1}. In both cases, the inelastic breakup
spectra are larger than the elastic ones. The elastic and inelastic spectra 
for neutron or proton both peak at about the same energy in both reactions 
with the proton spectra peaking at higher energy than the neutron ones.

The difference in the the maxima of the neutron and proton spectra in
Fig. \ref{fig1} can be 
interpreted in terms of the Coulomb deceleration of the deuteron and 
posterior Coulomb acceleration of the outgoing proton. At the radius at 
which breakup occurs, $R_{bu}$, we would expect the deuteron to have lost 
a kinetic energy of $Ze^2/R_{bu}$, which is later recovered by the outgoing
proton. On the average, we would thus expect to find for the outgoing
neutron and proton kinetic energies
\begin{eqnarray}
E_{n} & \approx & \frac{1}{2}\left(E_{d}-\frac{Ze^{2}}{R_{bu}}-\varepsilon_{d}\right)\\
E_{p} & \approx & \frac{1}{2}\left(E_{d}+\frac{Ze^{2}}{R_{bu}}-\varepsilon_{d}\right)\nonumber\,.
\end{eqnarray}
Interpreting the difference in the peaks as the energy 
difference at the most probable radius, we find the breakup radius $R_{bu}$ 
to be 10.6 fm for $^{48}$Ti and 14.6 fm for $^{181}$Ta. We compare these 
values to strong interaction radii of 6.0 fm for $^{48}$Ti and 8.6 fm for 
$^{181}$Ta, obtained by taking $R_{int}\approx 1.25(A^{1/3}+2^{1/3})$ fm, 
with $A$ the target mass number. We conclude that the breakup occurs 
predominantly in the Coulomb field of the target.

\begin{figure*}[b]
\centering
\resizebox{18pc}{!}{\includegraphics{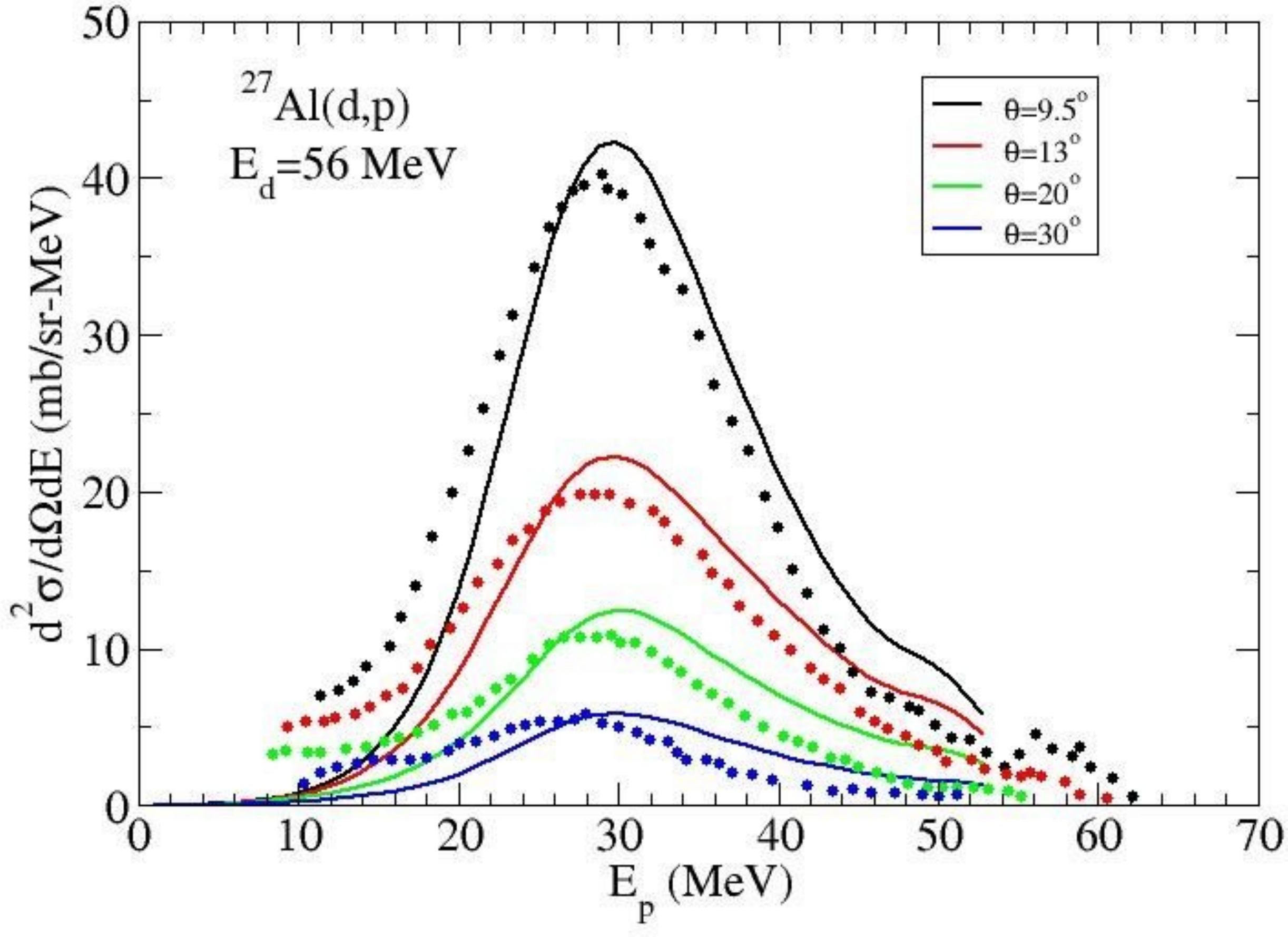}} 
\resizebox{18pc}{!}{\includegraphics{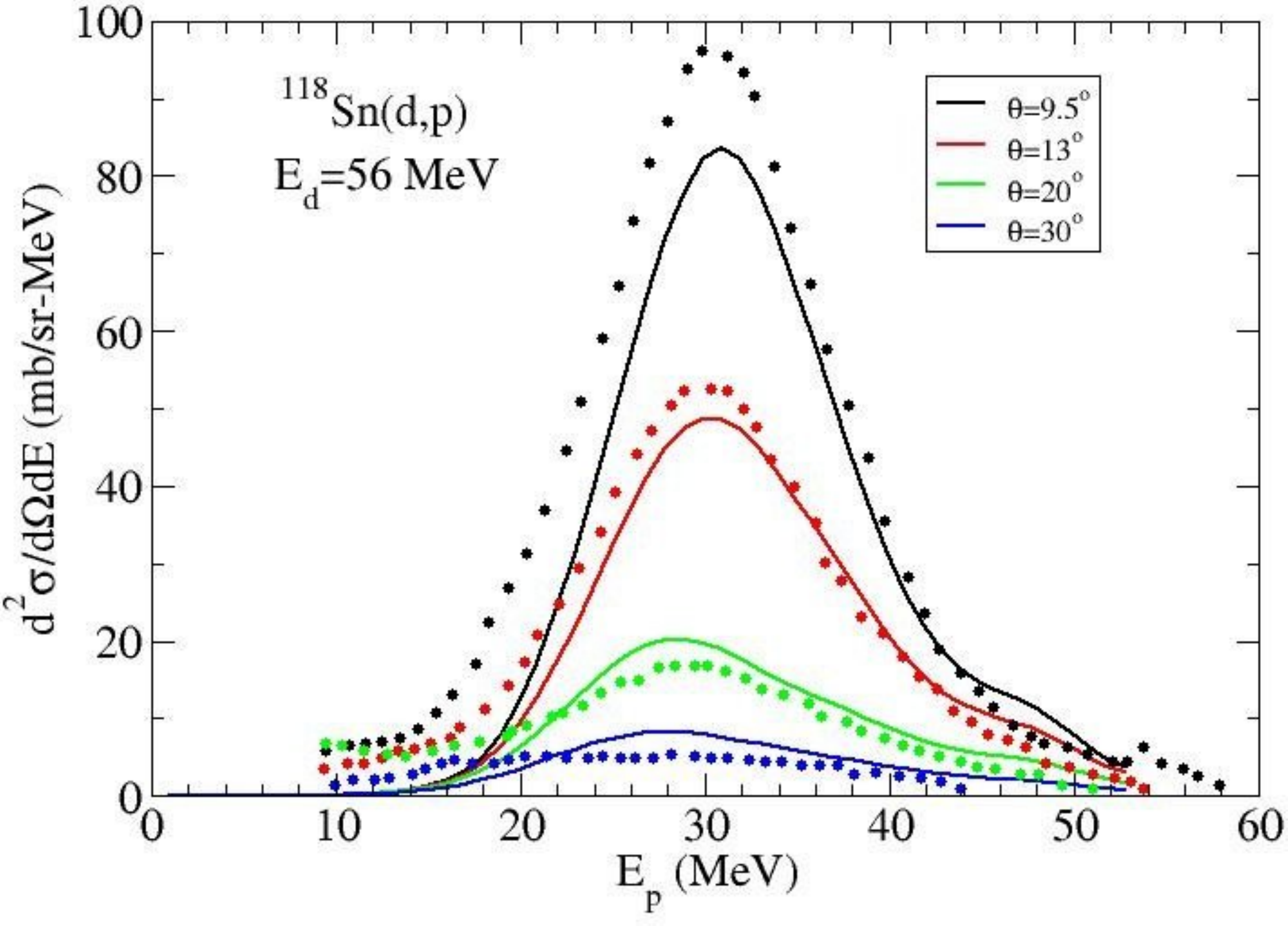}}
\caption{(a) Inclusive double differential proton spectra for 56
MeV deuterons incident on $^{27}$Al. (b) Inclusive double differential proton spectra for 56 MeV deuterons incident on $^{118}$Si.}
\label{fig2} 
\end{figure*}

In Fig. \ref{fig2}, we compare our calculations with the experimental inclusive double differential proton cross sections for 56 MeV deuterons incident on $^{27}$Al and $^{118}$Sn \cite{Matsuoka-80}. The cross sections include protons from both elastic and inelastic breakup.
The calculations agree well with the experimental data for the case of 
$^{27}$Al although the calculations are shifted to slightly higher energies, 
probably due to the fact that their center-of-mass motion is not extracted 
correctly. This shift is much smaller in the case 
of $^{118}$Sn. However, here the spectrum is underpredicted by about 20\% 
at an angle of 9.5$^{\circ}$ and by a few percent at 13$^{\circ}$. As the breakup 
at small scattering angles (corresponding to large impact parameters) is 
principally elastic, we suspect that more partial waves should be 
included in our elastic breakup calculation. 

\begin{figure*}[t]
\centering
\resizebox{18pc}{!}{\includegraphics{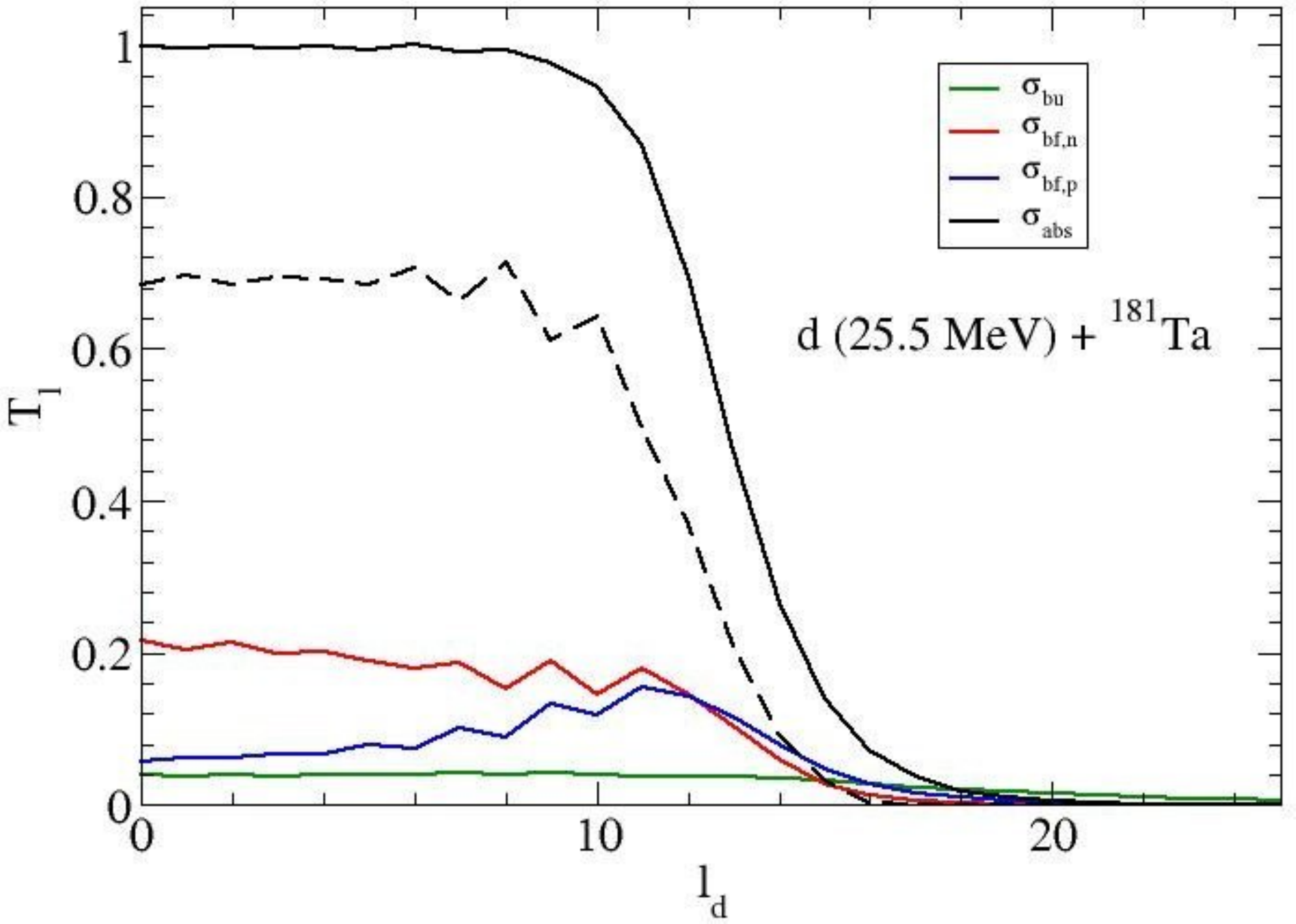}} 
\resizebox{18pc}{!}{\includegraphics{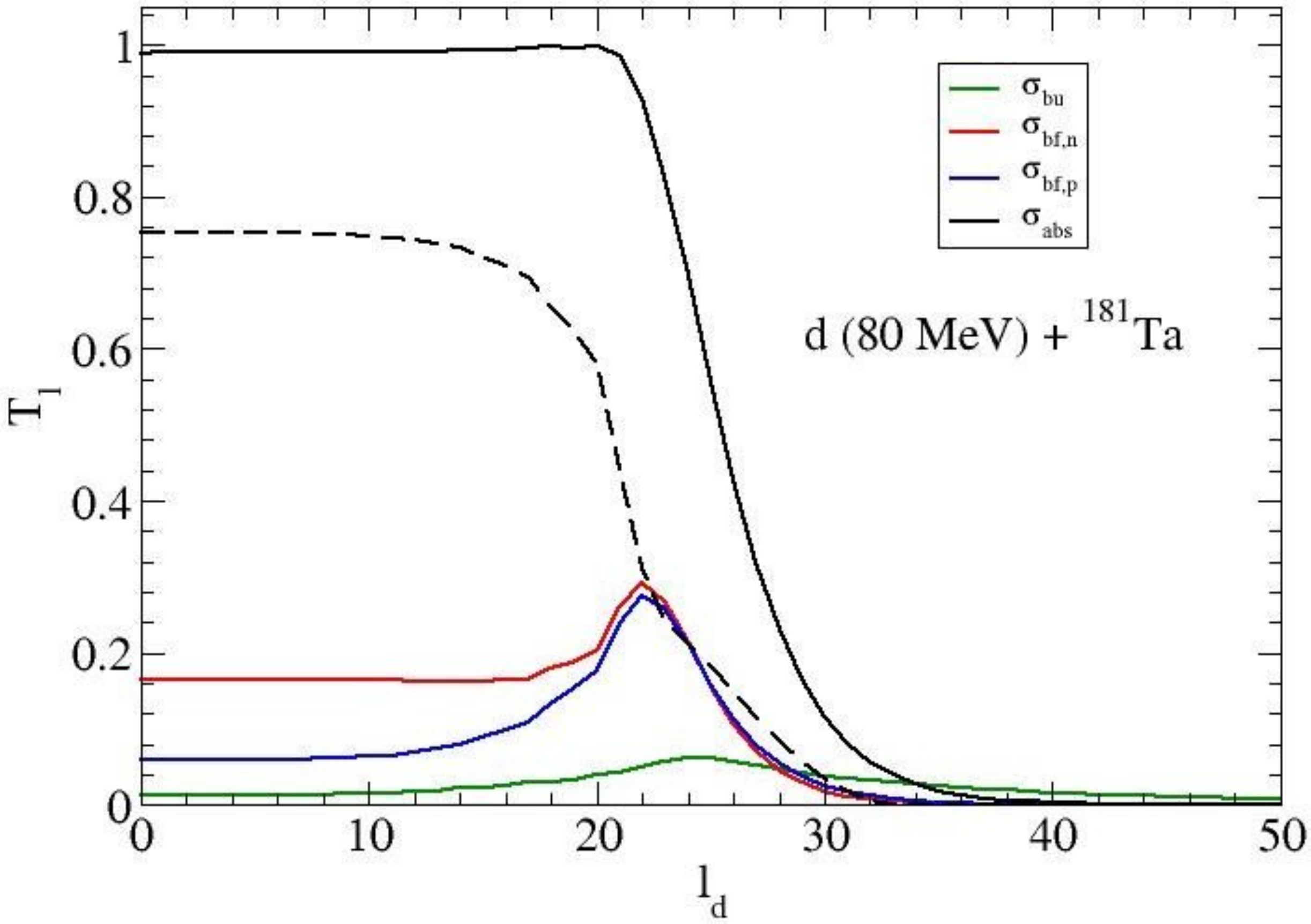}}
\caption{(a) Reaction probabilities as a function of the deuteron 
angular momentum for 25.5 MeV deuterons incident on $^{48}$Ti. 
(b) Reaction probabilities as a function of the deuteron angular 
momentum for 80 MeV deuterons incident on $^{181}$Ta.}
\label{fig3} 
\end{figure*}

A standard optical model calculation of deuteron scattering furnishes a
deuteron absorption probability (transmission coefficient) close to 1
below the grazing value of the angular momentum, corresponding to 
the solid black lines shown in Fig. \ref{fig3} for deuterons incident
at 25.5 and 80 MeV on $^{181}$Ta. 
The flux lost due to elastic and inelastic breakup reduces this probability
to about 75\% for low partial waves in the nuclear interior
and eliminates it in the surface region, as shown by the dashed lines in 
Fig. \ref{fig3}. Also shown are the neutron and proton
absorption probabilities from inelastic breakup, which 
dominate in the surface region and extend into the nuclear interior with
probabilities of about 10\% and 20\%, respectively. Only
elastic breakup extends to higher angular momenta outside the range
of the nuclear interaction.

\begin{figure*}[b]
\centering
\resizebox{18pc}{!}{\includegraphics{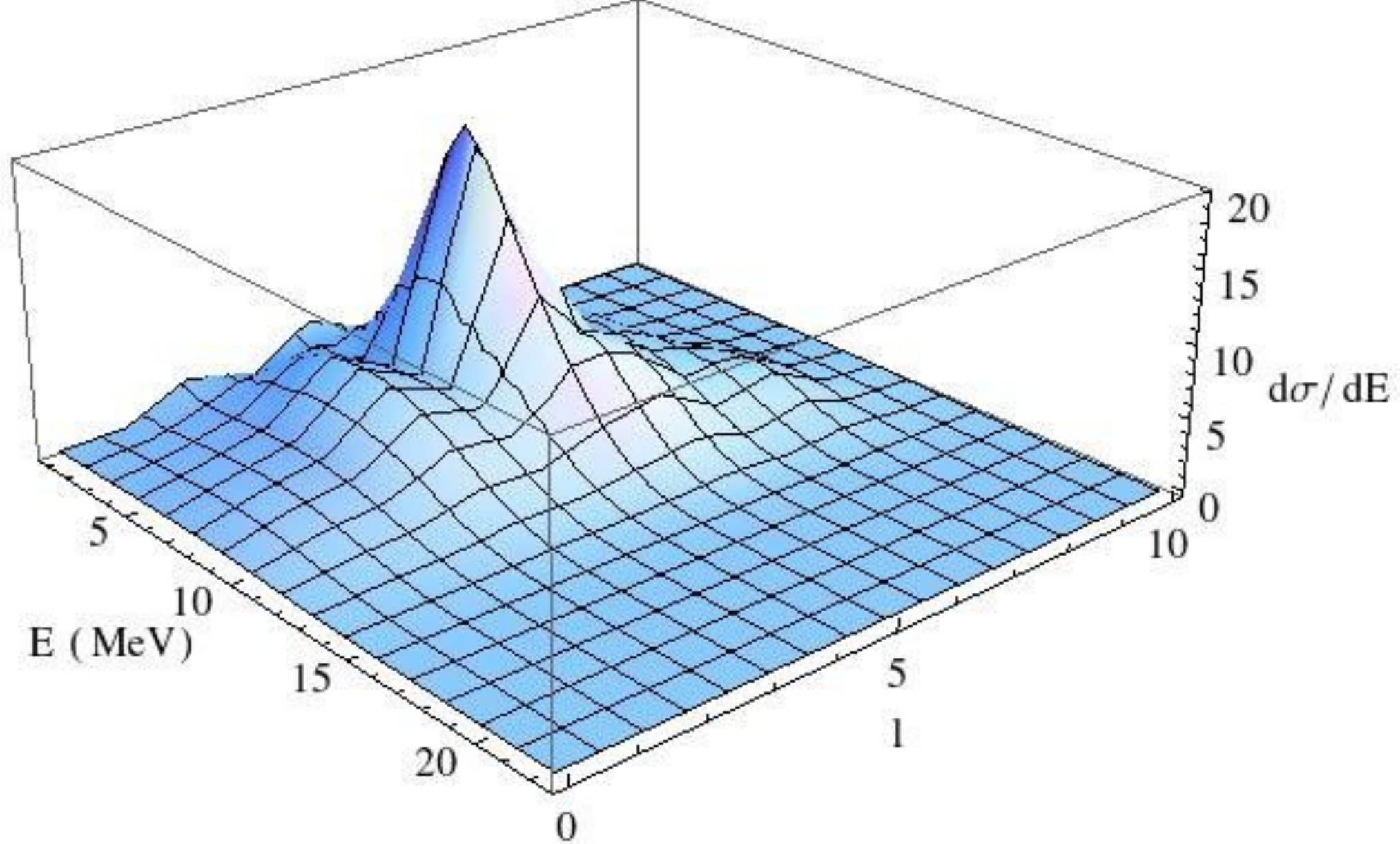}} 
\resizebox{18pc}{!}{\includegraphics{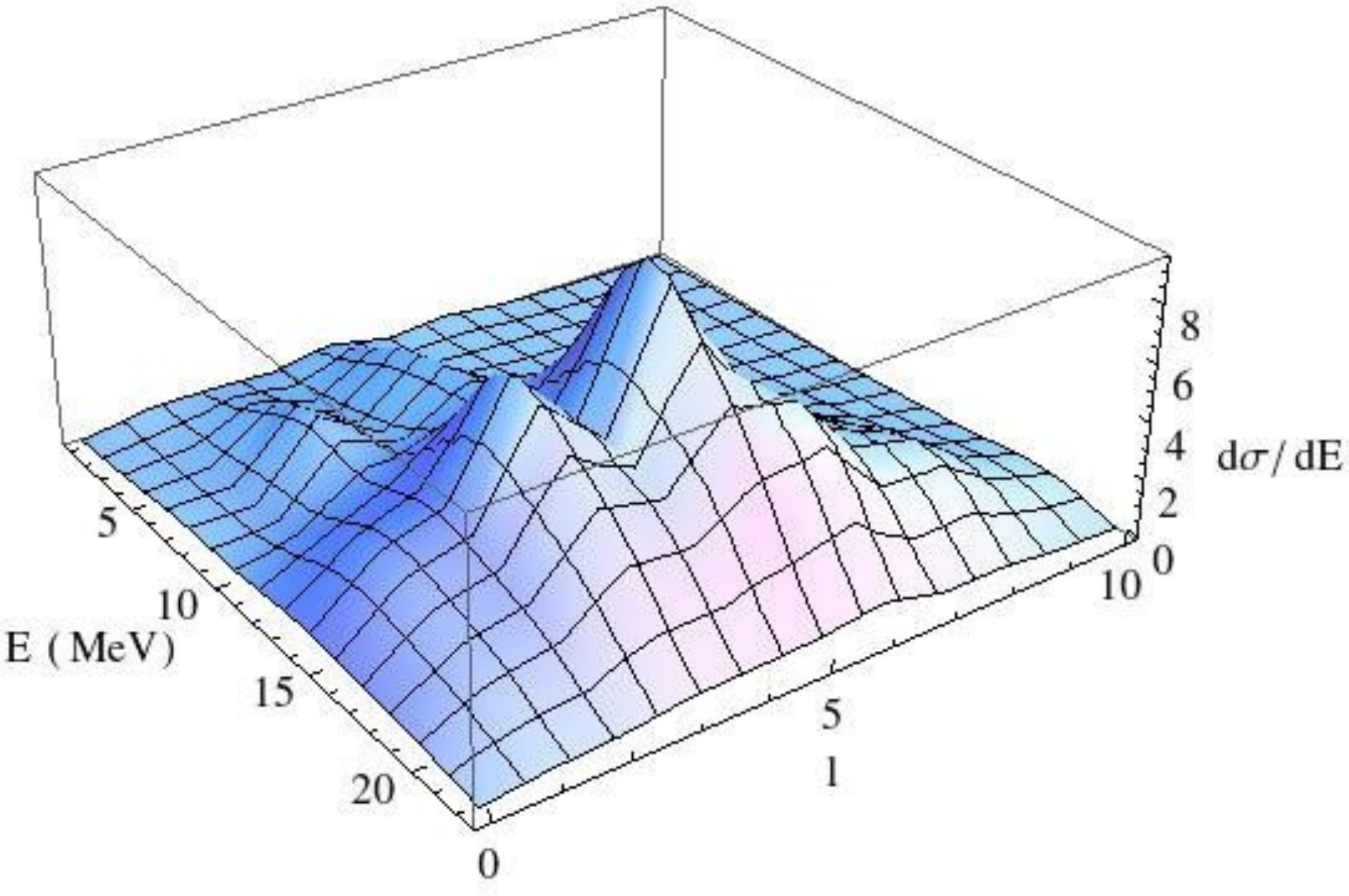}}
\caption{(a) Energy - angular momentum distribution of neutron + target 
differential CN formation cross section for 25.5 MeV deuterons 
incident on $^{181}$Ta. 
(b) Energy - angular momentum distribution of proton + target differential 
CN cross section formation for 25.5 MeV deuterons incident 
on $^{181}$Ta.}
\label{fig4} 
\end{figure*}

\begin{figure*}[t]
\centering
\resizebox{18pc}{!}{\includegraphics{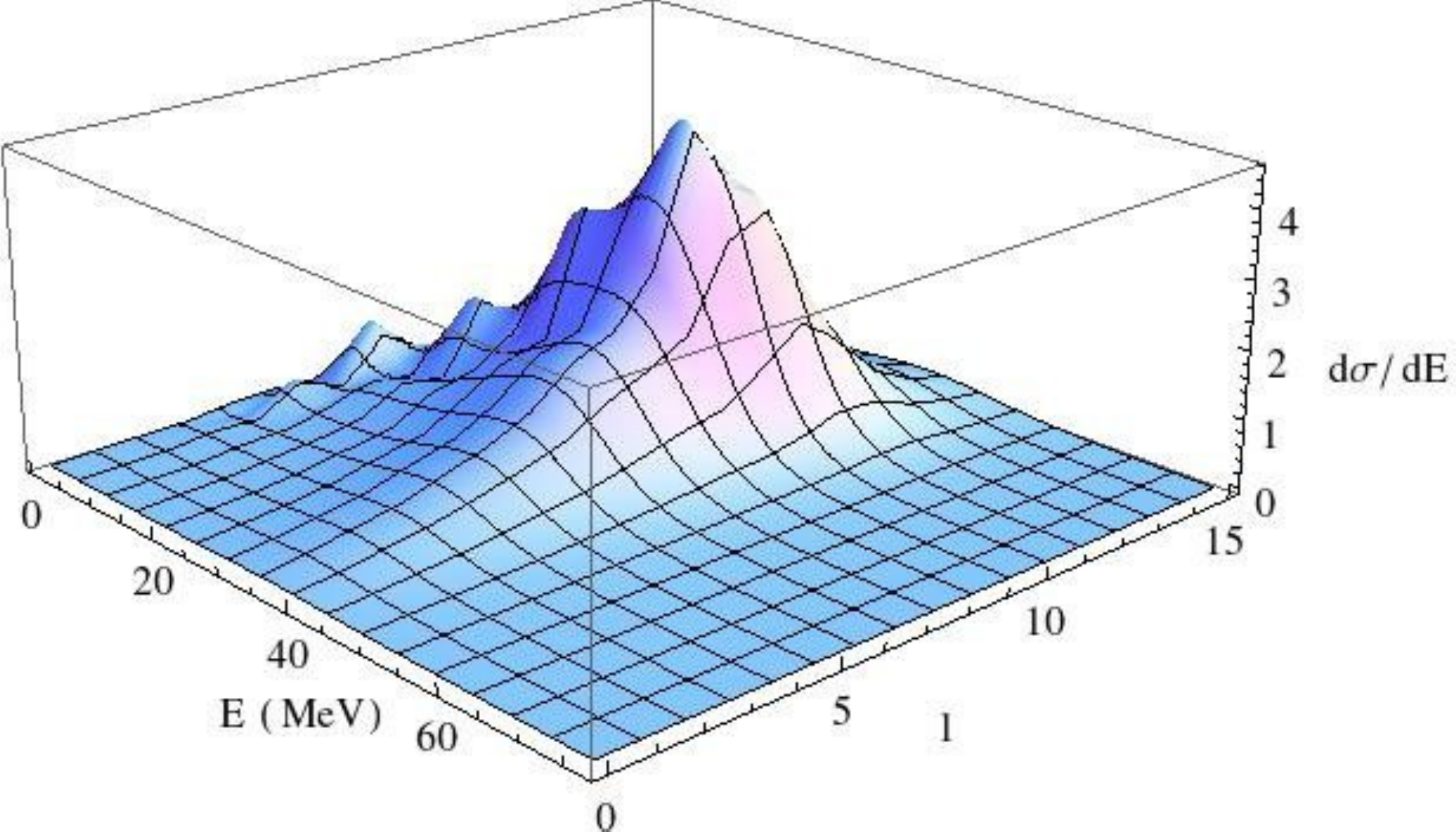}} 
\resizebox{18pc}{!}{\includegraphics{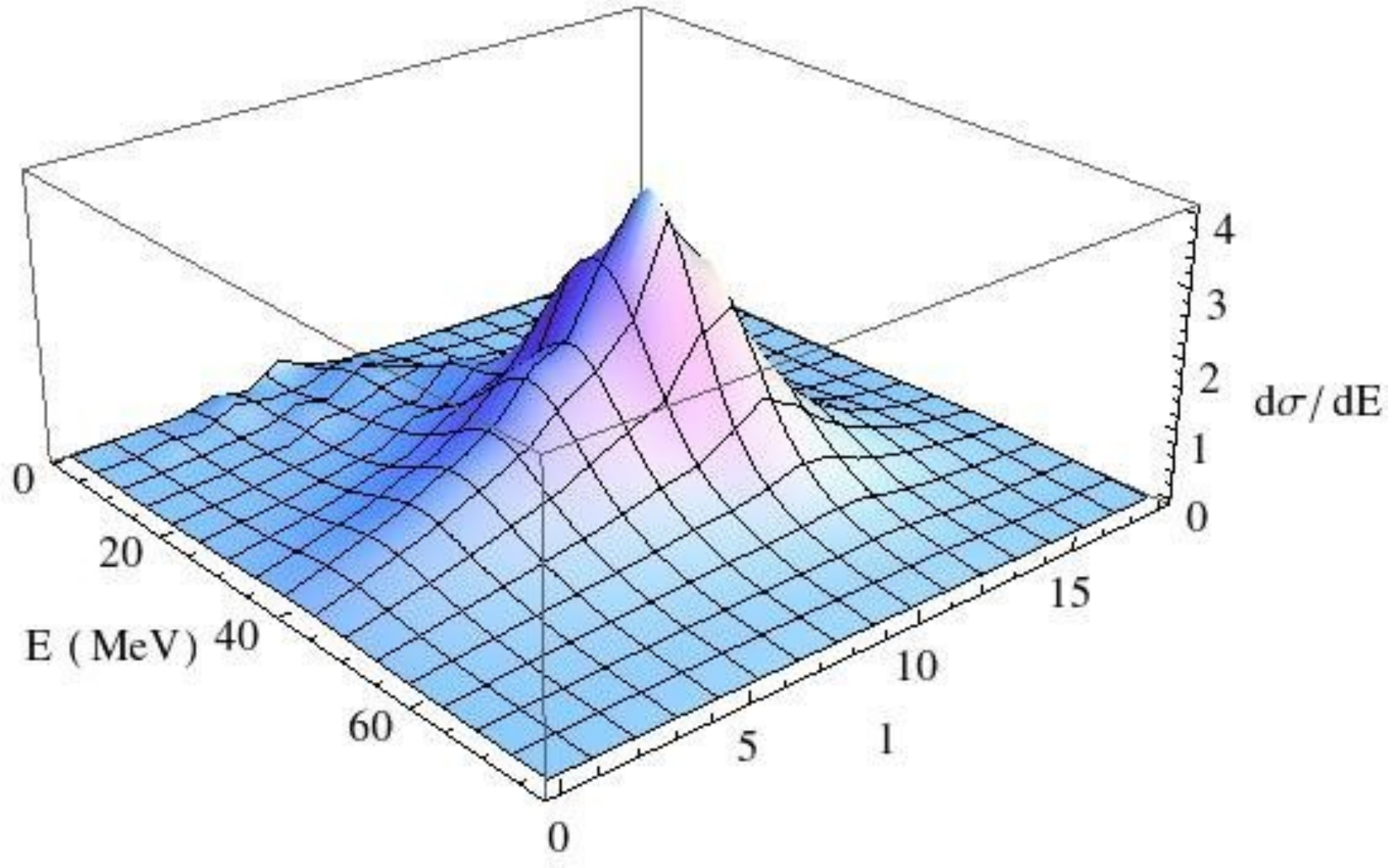}}
\caption{(a) Energy - angular momentum distribution of neutron + target
differential CN formation cross section for 80 MeV deuterons
incident on $^{181}$Ta. 
(b) Energy - angular momentum distribution of proton + target differential
CN cross section formation for 80 MeV deuterons incident on
$^{181}$Ta.}
\label{fig5} 
\end{figure*}

A complete calculation of a deuteron-induced reaction must also take into
account the formation and decay of the compound nucleus (CN) formed by
absorption of the deuteron, as well as those formed by aborption of the
neutrons or protons of inelastic breakup. The deuteron absorption cross
section is determined by the subtracted transmission coefficients, shown
as dashed lines in Fig. \ref{fig3}. The excitation energy of the CN
that results is determined by the kinetic energy of the incident deuteron.
Inelastic breakup reactions produce protons and neutrons over the entire
kinematically allowed range of energy. The corresponding differential 
formation cross sections are thus distributions in energy and angular
momentum. Two examples of these, for neutron and proton absorption due to
inelastic breakup in deuteron-induced reactions at incident energies of
25.5 MeV and 80 MeV, respectively, are shown in Figs. \ref{fig4} and \ref{fig5}.

In the reaction at 25.5 MeV, we observe in Fig. \ref{fig4} that the
neutron + target CN formation is well concentrated at lower energies while
the proton + target CN formation is concentrated for the most part at higher
energies, due to the importance of the Coulomb repulsion at this energy. The
separation in energy of the neutron and proton distributions is still visible
but much less distinct at 80 MeV, as can be seen in Fig. \ref{fig5}. At both
incident energies, the proton + target distribution extends to larger values
of the angular momentum, mainly due to the higher average energies of the
protons that are absorbed.

Well-defined structures in energy and angular momentum appear in the
differential cross sections at the lower incident deuteron energy of 25
Mev and, to a lesser extent, in the distributions at 80 MeV. These reactions
should thus be used with care as surrogates to other reactions \cite{Escher-12}.
The compound nucleus formed could have an initial energy-angular momentum distribution
very different from the one expected of a neutron- or proton-induced reaction. 

\section{Summary}

\label{sec:3}

We have used the post-form DWBA to calculate elastic and inelastic 
deuteron breakup cross sections in the zero-range limit. The breakup occurs
for the most part outside the range of the nuclear interaction. However,
the inelastic breakup cross sections, in which either the proton or neutron
is absorbed by the target, are larger than the elastic one, in which the
neutron and proton are simultaneously emitted. 

The breakup reactions reduce deuteron absorption at small impact parameters
by about 25\% and dominate the surface region completely. The
neutron + target absorption of inelastic breakup tends to occcur at lower
energy than the proton + target one, due to Coulomb repulsion,
which inhibits low-energy protons from entering the range of the nuclear
interaction. The low-energy inelastic breakup CN formation cross sections
are found to have well-defined structures in energy and angular momentum,
which could have important consequences in reactions in which they are
intended to serve as surrogates.

\begin{acknowledgements} BVC acknowledges partial
support from FAPESP (Project 2009/00069-5), the CNPq (Project 306692/2013-9) and
the International Atomic Energy Agency (Research Contract 17740).
\end{acknowledgements}

\end{document}